\documentclass[12pt]{article}%
\usepackage{amsmath,latexsym}
\usepackage{graphicx}
\usepackage{amsmath}
\usepackage{amsfonts}
\usepackage{amssymb}%
\begin{document}

\title{{Thin-shell wormholes from Kiselev
    black holes}}
   \author{
Peter K. F. Kuhfittig*\\
\footnote{E-mail: kuhfitti@msoe.edu}
 \small Department of Mathematics, Milwaukee School of
Engineering,\\
\small Milwaukee, Wisconsin 53202-3109, USA}

\date{}
 \maketitle

\begin{abstract}\noindent
This paper discusses the theoretical
construction of thin-shell wormholes from Kiselev
black holes.  We assume a barotropic equation
of state for the exotic matter on the shell.
While most of these wormholes are unstable
to linearized radial perturbations, a limit
argument is used to show that under certain
conditions, stable solutions can be found. \\


\end{abstract}

\section{Introduction}

A \emph{thin-shell wormhole}, first proposed by
Visser \cite{mV89}, is a theoretical construction
by the so-called cut-and-paste technique and
consists of grafting two black-hole spacetimes
together.  The junction surface is a three-dimensional
thin shell.  A wormhole can only be held open by
violating the null energy condition, requiring the
use of exotic matter \cite{MT88}.  For a thin-shell
wormhole, this requirement becomes less problematical
with the realization that the exotic matter is
confined to the thin shell.

One of the key issues in the study of thin-shell
wormholes is the stability of such structures to
linearized radial perturbations.  Once again,
Visser led the way by considering thin-shell
wormholes from Schwarzschild spacetimes using an
analysis based on the speed of sound.  It was
concluded that such wormholes are unstable  if
the speed of sound on the thin shell is less than
the speed of light.  (Due to the presence of exotic
matter, the speed of sound is presumably uncertain.)
Using Visser's approach, regions of stability have
been found for thin-shell wormholes from regular
charged black holes \cite{RK09} and from charged
black holes in generalized dilaton-axion gravity
\cite{RK10}.

Given that the exotic matter is confined to this
shell, a natural alternative to the above approach
is to assign an equation of state (EoS) to the
exotic matter.  Examples are the EoS of Chaplygin
and generalized Chaplygin gas \cite {ES07, BBC09,
eE09, SA13}.  A simpler, and in some ways, more
natural choice is the barotropic EoS $\mathcal{P}
=\omega\sigma$, where $\mathcal{P}$ is the surface
pressure and $\sigma$ the surface density.  Not
only is this the analogue of the EoS of a perfect
fluid, we will see in the next section that in
the present study, $\mathcal{P}=\omega\sigma$
may be the only acceptable EoS.

For the EoS $\mathcal{P}=\omega\sigma$, stable
solutions have proved to be fairly rare.  Thus,
in the discussion of thin-shell wormholes in
several spacetimes, Kuhfittig \cite{pK10} found
only unstable solutions for Schwarzschild, regular
charged, dilaton, and dilaton-axion spacetimes.
Other unstable solutions were found for Bardeen
and Hayward  thin-shell wormholes, respectively 
\cite{SJ16, pK}.

In this paper we study thin-shell wormholes from
Kiselev black holes.  As in most of the above cases,
such wormholes tend to be unstable.  Exceptions do
exist, however: it is shown by means of a limit
argument that for certain choices of the
parameters, stable solutions can be constructed.

\section{The Kiselev black hole}\label{S:Kiselev}

A Kiselev black hole is a spherically symmetric
black hole surrounded by quintessence dark
energy and is given by \cite{vK03}
\begin{equation}\label{E:line1}
ds^2 = -f(r) dt^2 + [f(r)]^{-1}dr^2
  + r^2(d\theta^2+\sin^2\theta\, d\phi^2),
\end{equation}
where
\begin{equation}\label{E:Kiselev}
    f(r)=1-\frac{2M}{r}-
    \frac{\alpha}{r^{3\omega_1+1}}.
\end{equation}
Here $\omega_1=p/\rho$, $\omega_1<-1/3$,
is the EoS of quintessence, $\alpha$
will be referred to as the Kiselev
parameter, and $M$ is the mass of the
black hole viewed from a distance.  (We
are using units in which $c=G=1$.)  Ref.
\cite{JY17} concentrates on the special
case $\omega_1=-2/3$, leading to
\begin{equation}\label{E:special}
    f(r)=1-\frac{2M}{r}-\alpha r.
\end{equation}
(This case is also considered below.)  By
setting $f(r)=0$, we obtain
\begin{equation*}
  r_{\pm}=\frac{1\pm\sqrt{1-8M\alpha}}
  {2\alpha},\quad
  0<\alpha \le\frac{1}{8M};
\end{equation*}
$r_{-}$ is the inner horizon and $r_{+}$
is the outer (cosmological) horizon.  The
special case $\alpha =1/8M$ yields a
Schwarzschild black hole with only one
horizon $r_h=4M$.

We can see from Eq. (\ref{E:special}) that
there is a curvature singularity at $r=0$.
On the other hand, both $r_{-}$ and $r_{+}$
are coordinate singularities, but if
$\alpha >1/8M$, then $f(r)=0$ has no real
solutions and we get a naked singularity.

It should be noted at this point that since
the Kiselev black hole assumes a quintessence
dark-energy background, the only plausible
EoS on the thin shell is the barotropic EoS
$\mathcal{P}=\omega_2\sigma$.

\section{Thin-shell wormhole construction}\noindent
As in Ref. \cite{PV95}, our construction begins
with two copis of a black-hole spacetime, Eq.
(\ref{E:line1}), and removing from each the
four-dimensional region
\begin{equation}\label{E:remove}
  \Omega^\pm = \{r\leq a\,|\,a>r_h\},
\end{equation}
where $r=r_h$ is the event horizon of the black
hole.  Now identify the time-like hypersurfaces
\begin{equation}
  \partial\Omega^\pm =\{r=a\,|\,a>r_h\}.
\end{equation}
The resulting manifold is geodesically complete and
possesses two asymptotically flat regions connected
by a throat.  Next, we use the Lanczos equations
\cite{mV89, eE09, PV95, LC04, ER04, TSE06, RKC06, RKC07,
RS07, LL08}
\begin{equation}\label{E:Lanczos}
  S^i_{\phantom{i}j}=-\frac{1}{8\pi}\left([K^i_{\phantom{i}j}]
   -\delta^i_{\phantom{i}j}[K]\right),
\end{equation}
where $[K_{ij}]=K^{+}_{ij}-K^{-}_{ij}$ and $[K]$ is the
trace of $K^i_{\phantom{i}j}$.  In terms of the surface
energy-density $\sigma$ and the surface pressure
$\mathcal{P}$, $S^i_{\phantom{i}j}=\text{diag}(-\sigma,
\mathcal{P}, \mathcal{P})$.  The Lanczos equations now
imply that
\begin{equation}\label{E:sigma}
  \sigma=-\frac{1}{4\pi}[K^\theta_{\phantom{\theta}\theta}]
\end{equation}
and
\begin{equation}\label{E:LanczosP}
  \mathcal{P}=\frac{1}{8\pi}\left([K^\tau_{\phantom{\tau}\tau}]
    +[K^\theta_{\phantom{\theta}\theta}]\right).
\end{equation}

According to Ref. \cite{PV95}, a dynamic analysis can be
obtained by letting the radius $r=a$ be a function of
proper time $\tau$.  The result is
\begin{equation}\label{E:sigma1}
\sigma = - \frac{1}{2\pi a}\sqrt{f(a) + \dot{a}^2}
\end{equation}
and
\begin{equation}\label{E:P}
  \mathcal{P} =  -\frac{1}{2}\sigma + \frac{1}{8\pi
}\frac{2\ddot{a} + f^\prime(a) }{\sqrt{f(a) + \dot{a}^2}},
\end{equation}
where the overdot denotes the derivative with respect to
$\tau$.  Since $\sigma$ is negative on this shell, we are
dealing with exotic matter.  (The reason is that the
radial pressure $p$ is zero for a thin shell; so
for the radial outgoing null vector $(1,1,0,0)$, we
have $T_{\alpha\beta}\mu^{\alpha}\mu^{\beta}
=\rho+p=\sigma +0<0$.)

Since $r=a$ is a function of time, one can obtain the
following relationship:
\[
   \frac{d}{d\tau}(\sigma a^2)+\mathcal{P}\frac{d}{d\tau}(a^2)=0.
\]
This equation can also be written in the form
\begin{equation}\label{E:conservation}
  \frac{d\sigma}{da} + \frac{2}{a}(\sigma+\mathcal{P}) = 0.
               \end{equation}
Still following Ref. \cite{PV95}, for a static
configuration of radius $a_0$, we have $\dot{a}=0$ and
$\ddot{a}=0$.  The ``linearized fluctuations around
a static solution," discussed below, are characterized
by the constants $a_0$ and $\sigma_0$.  Given the EoS
$\mathcal{P}=\omega_2\sigma$,
Eq. (\ref{E:conservation}) can be solved by separation of
variables and yields
\begin{equation}
  |\sigma(a)|=|\sigma_0|\left(\frac{a_0}{a}
    \right)^{2(\omega_2+1)},
\end{equation}
where $\sigma_0=\sigma(a_0)$.  So the solution is
\begin{equation}\label{E:sigmaexplicit}
  \sigma(a)=\sigma_0\left(\frac{a_0}{a}
  \right)^{2(\omega_2+1)},\quad \sigma_0=\sigma(a_0).
\end{equation}

Next, we rearrange Eq. (\ref{E:sigma1}) to obtain the
equation of motion
\begin{equation*}
\dot{a}^2 + V(a)= 0.
\end{equation*}
Here the potential $V(a)$ is defined  as
\begin{equation}\label{E:Vdefined}
V(a) =  f(a) - \left[2\pi a \sigma(a)\right]^2.
\end{equation}
Expanding $V(a)$ around $a_0$, we obtain
\begin{eqnarray}
V(a) &=&  V(a_0) + V^\prime(a_0) ( a - a_0) +
\frac{1}{2} V^{\prime\prime}(a_0) ( a - a_0)^2  \nonumber \\
&\;& + O\left[( a - a_0)^3\right].
\end{eqnarray}
Since we are linearizing around $a=a_0$, we require that
$V(a_0)=0$ and $V'(a_0)=0$.  The configuration is in
stable equilibrium if $V''(a_0)>0$.

For the Kiselev black hole, where
\[
   f(r)=1-\frac{2M}{r}-
   \frac{\alpha}{r^{3\omega_1+1}},
\]
we first obtain
\begin{equation}\label{E:sigma2}
   \sigma_0=-\frac{1}{2\pi a_0}
   \sqrt{1-\frac{2M}{a_0}-
   \frac{\alpha}{a_0^{3\omega_1+1}}}
\end{equation}
by Eq. (\ref{E:sigma1}) with $\dot{a}=0$.  Then
from Eq. (\ref{E:Vdefined}), and making use of
Eq. (\ref{E:sigmaexplicit}),
\begin{multline}\label{E:V}
  V(a)=1-\frac{2M}{a}-\frac{\alpha}
  {a^{3\omega_1+1}}-4\pi^2a^2\sigma^2\\
   =1-\frac{2M}{a}-\frac{\alpha}
  {a^{3\omega_1+1}}
     -4\pi^2a^2\sigma_0^2
     \left(\frac{a_0}{a}\right)^{4+4\omega_2}\\
     =1-\frac{2M}{a}-\frac{\alpha}
     {a^{3\omega_1+1}}
     -\left(1-\frac{2M}{a_0}-\frac{\alpha}
  {a_0^{3\omega_1+1}}\right)
  \frac{a_0^{2+4\omega_2}}{a^{4+4\omega_2}}.
\end{multline}
Observe that $V(a_0)=0$.  To satisfy the
condition $V'(a_0)=0$, we need to compute
$\omega_2$.  From
\begin{equation}\label{E:Vprime}
   V'(a_0)=\frac{2M}{a_0}+\alpha
   (3\omega_1+1)\frac{1}{a_0^{3\omega_1+1}}
   +\left(1-\frac{2M}{a_0}-
   \frac{\alpha}{a_0^{3\omega_1+1}}
   \right)(2+4\omega_2)=0,
\end{equation}
we get
\begin{equation}\label{E:omega2}
   \omega_2=\frac{1}{4}
   \frac{-2+\frac{2M}{a_0}+\alpha
      \frac{1-3\omega_1}{a_0^{3\omega_1+1}}}
      {1-\frac{2M}{a_0}-\frac{\alpha}
      {a_0^{3\omega_1+1}}}.
\end{equation}

To allow a substitution into the second
derivative, we need to write $V''(a_0)$ in the
following convenient form:
\begin{multline}\label{E:secder1}
  V''(a_0)=\frac{2}{a_0^2}\left[-\frac{2M}{a_0}
  -\frac{1}{2}\alpha (3\omega_1+1)(3\omega_1+2)
      a_0^{-3\omega_1-1}\right.\\
   \left. -\left(1-\frac{2M}{a_0}
   -\frac{\alpha}{a_0^{3\omega_1+1}}
   \right)
   (1+2\omega_2)(3+4\omega_2)\right].
\end{multline}
As an intermediate step, let us compute
\begin{equation*}
   1+2\omega_2=-\frac{1}{2}
   \frac{\frac{2M}{a_0}+\alpha
   \frac{3\omega_1+1}{a_0^{3\omega_1+1}}}
     {1-\frac{2M}{a_0}
     -\frac{\alpha}{a_0^{3\omega_1+1}}}
\end{equation*}
and
\begin{equation*}
   3+4\omega_2=\frac{1-\frac{4M}{a_0}-\alpha
   \frac{3\omega_1+2}{a_0^{3\omega_1+1}}}
   {1-\frac{2M}{a_0}
     -\frac{\alpha}{a_0^{3\omega_1+1}}}.
\end{equation*}
and substitute in Eq. (\ref{E:secder1}).  After
simplifying, we obtain
\begin{multline}\label{E:secder2}
  V''(a)|_{a=a_0}=\frac{2}{a^2}
\left[-\frac{2M}{a}-\frac{1}{2}\alpha
   (3\omega_1+1)(3\omega_1+2)
       \frac{1}{a^{3\omega_1+1}}
       \phantom{\frac{\frac{1+2^2}{1+M^2}}
          {\frac{1+3^2}{1+a^2}}}\right.\\
       \left. +\left(\frac{M}{a}
       +\frac{1}{2}\alpha\frac{3\omega_1+1}
       {a^{3\omega_1+1}}\right)
       \frac{1-\frac{4M}{a}-\alpha
       \frac{3\omega_1+2}{a^{3\omega_1+1}}}
       {1-\frac{2M}{a}-
       \frac{\alpha}{a^{3\omega_1+1}}}
       \right]_{a=a_0}.
\end{multline}
Observe that $V''$ is continuous for $a>0$.

\section{Unstable solutions}
Referring to Eq. (\ref{E:secder2}), the special case
$\omega_1=-2/3$ leads to
\begin{equation}\label{E:secder3}
   V''(a_0)=V''(a)|_{a=a_0}=\frac{2}{a^2}\left[
   -\frac{2M}{a}+\frac{\left(\frac{M}{a}-\frac{1}{2}
   \alpha a\right)\left(1-\frac{4M}{a}\right)}
   {1-\frac{2M}{a}-\alpha a}\right]_{a=a_0}.
\end{equation}
This case is discussed in Refs. \cite{JY17} and
\cite{YJBH} and will be of interest to us, as
well.  For now it is sufficient to observe that
the choice $\omega_1=-2/3$ leads to an unstable
solution: suppose $M=1$ and $\alpha =1/16$,
producing the usual two horizons characteristic
of the Kiselev black hole.  The graphs of
$V''(a)$ and $f(a)=1-2/a-\alpha a$ are shown
\begin{figure}[tbp]
\begin{center}
\includegraphics[width=0.8\textwidth]{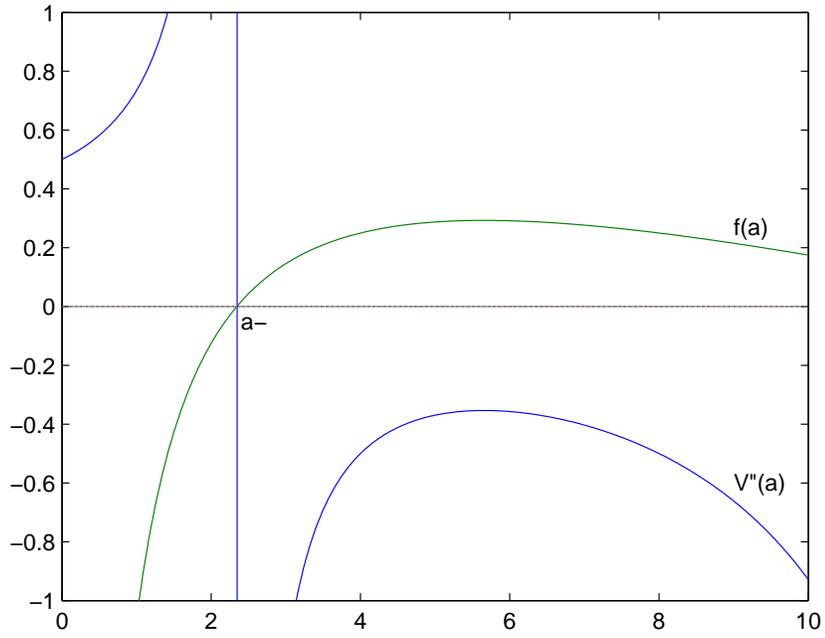}
\end{center}
\caption{The wormhole is unstable whenever
   $a_0>a_-$.}
\end{figure}
in Fig. 1.  It is clear from the figure that
$V''(a_0)<0$ whenever $a_0>a_-$, the inner
horizon.

\begin{figure}[tbp]
\begin{center}
\includegraphics[width=0.8\textwidth]{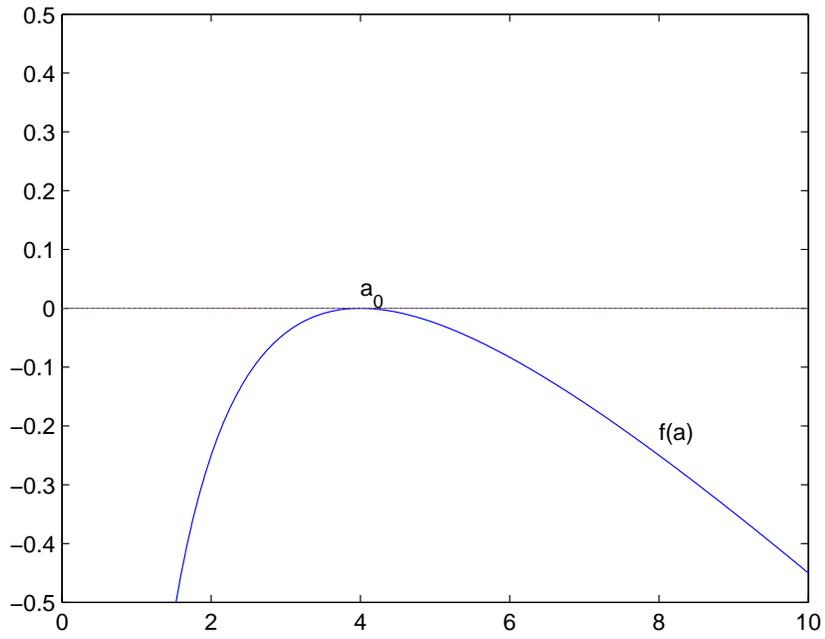}
\end{center}
\caption{There is a single horizon at $a=a_0$.}
\end{figure}

\section{Stable solutions}\label{S:ss}
First we recall that for a Kiselev black hole, the
Kiselev parameter is restricted to $0<\alpha\le
1/8M$.  So if $\alpha =1/8M$ and $M=1$, we obtain
from Eqs. (\ref{E:Kiselev}) and (\ref{E:special}),
\begin{equation}
   f(a)=1-\frac{2}{a}-\frac{a}{8},
\end{equation}
yielding a single horizon at some $a=a_0$, shown
in Fig. 2.  For $\alpha >1/8M$, we obtain a naked
singularity.  This case yields a stable solution.
(Stable solutions from naked singularities are
also discussed Refs. \cite{eE09} and \cite{pK10}.)
To see why, we choose $M=1$ again.  Then for
an arbitrarily small $\eta >0$, we obtain a new
continuous function of $a$ and $\eta$:
\begin{equation}
   f(a, \eta)=1-\frac{2}{a}-\frac{a}{8-\eta}<0.
\end{equation}
So
\begin{equation}\label{E:single}
\text{lim}_{\eta\rightarrow 0}f(a_0,\eta)
   =f(a_0,0)=f(a_0)=0.
\end{equation}
Substituting in Eq. (\ref{E:secder3}) with the
same $\alpha$ yields another new function, this
time denoted by
\begin{equation}
   V''(a, \eta)=\frac{2}{a^2}\left[-\frac{2}{a} +
   \frac{\left(\frac{1}{a}-\frac{a}{2(8-\eta)}
    \right)\left(1-\frac{4}{a}\right)}
      {1-\frac{2}{a}-\frac{a}{8-\eta}}\right].
\end{equation}
The graph of $V''(a, \eta)$, together with
$f(a, \eta)$, in Fig. 3 shows two stable regions
where $V''(a, \eta)>0$.  Observe that
$V''(a, \eta)$ is a continuous function of $a$
and $\eta$.  In particular,
\begin{figure}[tbp]
\begin{center}
\includegraphics[width=0.8\textwidth]{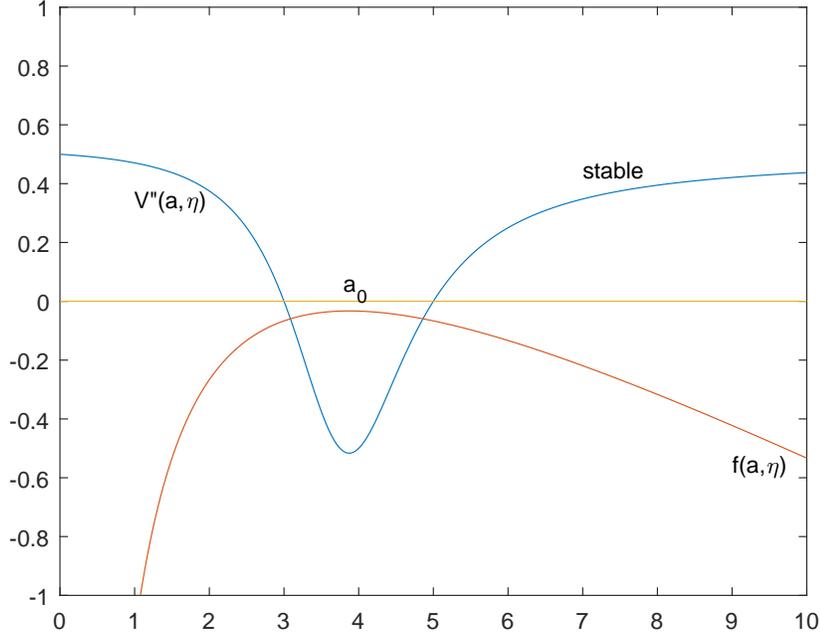}
\end{center}
\caption{Plot showing the region of stability
   where $V''(a,\eta)>0$.}
\end{figure}
\begin{equation}\label{E:Vlimit}
   \text{lim}_{\eta\rightarrow 0}
       V''(a, \eta)=V''(a, 0)=V''(a).
\end{equation}

To show that one can obtain a stable solution
without a naked singularity, we will use a limit
argument that depends on the continuity of
$V''$. Returning to Fig. 3, the two stable regions
on opposite sides of $a_0$ begin at some value
$a$, but since $\eta$ is arbitrarily small, $a$
can be placed as close to $a_0$ as we please.
To illustrate this point, if $\eta =0.5$, then
$V''=0$ at $a=3$ and $a=5$.  This is the case
shown in Fig. 3.  If $\eta =0.1$, then the zeros
of $V''$ are closer together at $a=3.55$ and
$a=4.45$, respectively, and if $\eta =0.00001$,
the respective zeros are at $a=3.996$ and
$a=4.004$, clearly showing the trend, a shrinking
interval centered at $a_0$.  Now we can simply
state that $V''>0$ outside the closed interval
$[a_0-a, \,a_0+a]$.  In particular, if we choose a
constant $\beta >0$, then $V''(a_0-a-\beta, \,\eta)>0$
and $V''(a_0+a+\beta, \,\eta)>0$ since every
$a_0\pm (a+\beta)$ lies outside the interval.  Next,
let us consider a positive sequence $\{a_n\}$ such
that $\text{lim}_{\rightarrow \infty}a_n=0$.
Then
\[
     [a_0-a_n, \,a_0+a_n]\rightarrow a_0, \,
     \text{to be denoted by} \, [a_0, \,a_0],
\]
while
\[
   [a_0-a_n-\beta, \,a_0+a_n+\beta]\rightarrow
   [a_0-\beta, \,a_0+\beta].
\]
From the condition
\begin{equation*}
   V''(a_0\pm (a_n+\beta, \,\eta))>0,
\end{equation*}
it now follows from the continuity of
$V''(a, \eta)$ that
\begin{equation*}
   \text{lim}_{n\rightarrow \infty,\,
   \eta\rightarrow 0}
   V''(a_0\pm (a_n+\beta),\,\eta)=
   V''(a_0\pm\beta,\,0)=V''(a_0\pm\beta),
\end{equation*}
making use of Eq. (\ref{E:Vlimit}).  But
$V''(a_0\pm\beta)>0$ since every
$a_0\pm\beta$ lies outside the interval
$[a_0,a_0]$.  Referring to Eq.
(\ref{E:single}), since $f(a_0,0)=f(a_0)=0$,
there is an event horizon at $a=a_0$ and
we obtain a stable solution without a
naked singularity.

We know from Sec. \ref{S:Kiselev} that the
special case $\alpha =1/8M$ of the Kiselev
parameter yields a Schwarzschild black hole
with only one event horizon, denoted by
$a_0$ in Fig. 2.  We have seen that the
limiting case still fits the Kiselev model
and therefore has a special significance:
the thin-shell wormhole obtained is stable
to linearized radial perturbations.  By
contrast, in Visser's original formulation
discussed in the Introduction, thin-shell
wormholes are unstable if constructed
directly from the traditional Schwarzschild
spacetime, by which is meant a black hole
that is not surrounded by quintessence
dark energy  [Eq. (\ref{E:Kiselev}) with
$\alpha =0$] unless the speed of sound
exceeds the speed of light.

As a final comment, the region of interest
in Fig. 3 is now restricted to the region on
the right side since we have to remain
outside the event horizon.

\section{Conclusion}
This paper employs the standard cut-and-paste
technique for the theoretical construction of
thin-shell wormholes from Kiselev black holes.
Since a Kiselev black hole is surrounded by
quintessence dark energy, the only reasonable
equation of state for the exotic matter on
the shell is the barotropic EoS $\mathcal{P}
=\omega\sigma$.  Most such wormholes turn out
to be unstable to linearized radial
perturbations.  Exceptions do exist, however:
by using the continuity of $V''$, it is
shown by means of a limit argument that for
certain choices of the parameters, stable
solutions exist due to the
quintessence-dark energy background.  The
results in this paper complement those in
Refs. \cite{RK09} and \cite{RK10}, which
also deal with stable thin-shell wormholes.

\end{document}